В.В. Кромер

г. Новосибирск

**СУММА ТЕСТОЛОГИИ**

0. Данные тезисы подготовлены для доклада на Международной научно-практической конференции "Инновации в педагогическом образовании" (Новосибирский государственный педагогический университет, 22–24 октября 2007 г).

1. В данном докладе сформулированы требования к педагогическим измерительным материалам (тестам), проанализированы применяющиеся измерительные материалы и причины, снижающие их качество. Рассмотрена возможность улучшения качества тестов.

2. Качество тестов (тестовых результатов) характеризуется валидностью и надежностью. Валидность теста зависит от валидности составляющих его заданий, а надежность – в основном от их количества. При предтестировании валидность заданий определяется измерением коэффициента корреляции заданий с критерием $r$. Недостаточный размер нормативной выборки – причина измерения коэффициента корреляции с недопустимой погрешностью, в результате чего в тест проникают некачественные задания. Пример: Преобразованные по формуле преобразования Фишера [1, с. 380] $z = \frac{1}{2}\ln\frac{1+r}{1-r} = x + \frac{x^3}{3} + \frac{x^5}{5} + \ldots$ значения распределены с дисперсией $\frac{1}{n-3}$, что при выборке в $n=100$ испытуемых дает доверительный интервал для $z$-значения с полушириной 0,2. Таким образом, задание с оцененной на нормативной выборке еще допустимой корреляцией $r = 0,30$ и включенное на этом основании в тест, может иметь совершенно неприемлемый истинный коэффициент корреляции $r = 0,11$, и т.п. Аналогичные выводы делаются и при использовании взамен коэффициента корреляции индекса дискриминативности. Выход заключается в применении моделей со снижением в процессе рабочего тестирования веса некачественных заданий на основе параметризации заданий [2; 3].

3. Низкое качество существующих тестов выявляется при оценке количеств различимых интервалов (квантов) в рабочем диапазоне измерения теста. Существует простое мнемоническое правило: качество существующих тестов оценивается отметкой на привычной четырехбалльной шкале, равной числу различимых тестом квантов измерения [6] (См. Приложение). Тем самым удовлетворительными признаются тесты, различающие хотя-бы три уровня измеряемого качества, чего совершенно недостаточно для дифференциации испытуемых в соответствии с заявленными целями тестирования.

4. Один из источников смещенности значений тестовых баллов – неучет явления угадывания. Возврат к несмещенным значениям тестовых баллов в заданиях с выбором одного правильного ответа возможен при введении коррекции баллов на угадывание, что требует раздельного учета факта выбора неправильного ответа и отказа от ответа (пропуска задания), либо же при комплектовании теста заданиями, где вероятность угадывания верного ответа мала (ниже 1%). Это задания открытой формы и задания с выбором нескольких ответов, на установление соответствия и на установление правильной последовательности при соответствующем выборе числа элементов в задании [5].

5. При тестировании параметры тестирующих определяются тестовыми заданиями, а параметры заданий – испытуемыми, т.е. речь идет о взаимно согласованной параметризации испытуемых и заданий. Если по результатам предтестирования из теста удаляются некачественные задания, то удалять из матрицы неадекватных испытуемых недопустимо. Выход заключается в переходе к моделям тестирования, где параметризируется степень неадекватности испытуемого, и данные испытуемые принимают участие в параметризации с очень малым весом [2; 3].

6. При извлечении из матриц данных параметров сторон тестирования в IRT-моделях возникает проблема испытуемых с экстремальными значениями тестовых баллов [7, с. 111–117]. Байесовский подход к проблеме, при всей корректности получаемых результатов, ведет к низкой очевидной валидности теста, а тем самым к отказу от байесовского подхода [4].

**Список литературы**

1. Варден Ван дер. Математическая статистика. – М.: ИИЛ, 1960.

2. Кромер В.В. Об одной возможности расширения семейства логистических моделей // Вопросы тестирования в образовании. – 2005. – № 3 (15). – С. 13–15.

3. Кромер В.В. О многопараметрической оценке уровней подготовленности испытуемых и трудностей заданий // Педагогические измерения. – 2005. № 3. – С. 65–72.

4. Кромер В.В. Добавление виртуальных заданий как альтернатива удалению реальных испытуемых // Вопросы тестирования в образовании. – 2005. – № 4(16). – С. 57–64.

5. Кромер В.В. Еще раз о коррекции тестового балла // Педагогические измерения. – 2007. – № 1. – С. 89–94.

6. Кромер В.В. Протестировали. С какой точностью? // Вестник педагогических инноваций. – 2007. – № 3(11). В печати.

7. Suen H.K. Principles of Test Theories. – Hillsdale, NJ: Erlbaum, 1990.

**Приложение**

В теории измерений вводится понятие об энтропийной погрешности $\varDelta_э$, линейно связанной со среднеквадратической погрешностью $s_e$ соотношением $\varDelta_э = k_э s_e$, где $k_э$ – энтропийный коэффициент, зависящий от вида распределения погрешности. Для равномерного распределения погрешности $k_э = \sqrt{3} \approx 1{,}73$, для нормального распределения $k_э = \sqrt{\dfrac{\pi e}{2}} \approx 2{,}07$. Энтропийное значение погрешности равно максимальной погрешности при прямоугольном законе распределения погрешности, эквивалентном с точки зрения вносимой дезинформации [Справочник по теоретическим основам радиоэлектроники. В двух томах. Том 2. Под ред. Б.Х. Кривицкого. М.: Энергия, 1977. С. 110].

При нормальном распределении тестовых баллов ожидаемое значение размаха тестовых баллов $\varDelta_z$ в единицах Z-шкалы (сигмах) зависит от объема тестируемой выборки $N$ и составляет 6 сигм при $N$ порядка 700, увеличиваясь примерно на 0,4 сигмы при каждом последующем удвоении выборки. Принимаем в среднем размах тестовых баллов в 7 сигм. При стандартной ошибке измерения тестового балла в $s_e$ эквивалентная ширина одного различимого интервала (кванта) в рабочем диапазоне измерительного инструмента (теста) определяется как $2\varDelta_э = 2k_э s_e \approx 4{,}14 s_e$, и при $s_e = \sqrt{1 - r_н}$ (сигма-единиц) число квантов составляет

$$n = \dfrac{\varDelta_z}{2\varDelta_э} = \dfrac{7}{4{,}14\sqrt{1 - r_н}} = \dfrac{1{,}69}{\sqrt{1 - r_н}}.$$

Сведем в таблицу результаты расчетов числа квантов измерения в зависимости от ожидаемой надежности теста. За основу соответствия между коэффициентом надежности и характеристикой надежности взяты данные из [Аванесов В.С. Основы научной организации педагогического контроля в высшей школе. М.: МИСиС, 1989. С. 148].

Таблица. Количество квантов измерения в зависимости от коэффициента надежности теста

| Коэффициент надежности $r_н$ | Характеристика надежности | Кол-во квантов |
|---|---|---|
| 0,99 | Практически не встречается | 16,9 |
| 0,90 | Отличная | 5,3 |
| 0,80 | Хорошая | 3,8 |
| 0,70 | Удовлетворительная | 3,1 |
| 0,50 | Неудовлетворительная | 2,4 |

Таким образом, удовлетворительными признаются тесты, различающие около трех интервалов в рабочем диапазоне измерения, а для отличных это число лишь несколько превышает пять, откуда вытекает простое мнемоническое правило: качество существующих тестов оценивается отметкой на привычной четырехбалльной шкале, равной числу различаемых ими квантов измерения. Напрашивается аналогия тестовой шкалы со шкалой бытового барометра, где при сотнях нанесенных шкальных отметок и высокой разрешающей способности, равной 1 мм рт. ст., реальная оценка делается по пяти не привязанным жестко к значениям давления лингвистическим переменным ЯСНО–ПЕРЕМЕННО–ПАСМУРНО–ДОЖДЬ–БУРЯ. Говорить о качественном педагогическом измерении можно лишь при применении тестов с коэффициентами надежности порядка 0,95 и выше, каковых очень мало. Считается, что в тестологической практике надежность тестов колеблется в интервале 0,8– 0,9 [Челышкова М.Б. Теория и практика конструирования педагогических тестов. – М.: Логос, 2002. С. 336]. Результаты исследования, приведенные в [Хлебников В.А., Овчинников В.В. О точности измерения тестового балла // Вопросы тестирования в образовании. 2002. № 4. С. 53–62], подтверждают сделанный вывод о невысокой точности используемых тестов.